\documentclass[runningheads]{llncs}

\usepackage[french]{babel}
\usepackage{graphicx}

\begin{document}

\titlerunning{\emph{Actes du symposium SSTIC06}}
\authorrunning{\emph{Actes du symposium SSTIC06}}

\title{Outrepasser les limites des techniques classiques de Prise d'Empreintes gr\^ace aux R\'eseaux de Neurones} 

\author{
Javier Burroni\inst{1} 
\and Carlos Sarraute\inst{2}
}
\institute{Equipe de d\'eveloppement d'Impact, Core Security Technologies.
\and Laboratoire de recherche de Core Security Technologies 
et D\'epartement de Math\'ematiques de l'Universit\'e de Buenos Aires.
}

\maketitle

\begin{abstract}
Nous pr\'esentons la d\'etection distante de syst\`emes d'exploitation comme 
un probl\`eme d'inf\'erence : \`a partir d'une s\'erie d'observations 
(les r\'eponses de la machine cible \`a un ensemble de tests), 
nous voulons inf\'erer le type de syst\`eme d'exploitation qui g\'en\`ererait ces observations 
avec une plus grande probabilit\'e. 
Les techniques classiques utilis\'ees pour r\'ealiser cette analyse pr\'esentent 
plusieurs limitations. 
Pour outrepasser ces limites, nous proposons l'utilisation de R\'eseaux de Neurones et d'outils statistiques. 
Nous pr\'esenterons deux modules fonctionnels : 
un module qui utilise les points finaux DCE-RPC pour distinguer les versions de Windows, 
et un module qui utilise les signatures de Nmap pour distinguer les versions de syst\`emes 
Windows, Linux, Solaris, OpenBSD, FreeBSD et NetBSD. 
Nous expliquerons les d\'etails de la topologie et du fonctionnement des r\'eseaux de neurones utilis\'es, 
et du r\'eglage fin de leurs param\`etres. 
Finalement nous montrerons des r\'esultats exp\'erimentaux positifs.
\end{abstract}

\section{Introduction}

Le probl\`eme de la d\'etection \`a distance du syst\`eme d'exploitation,
aussi appel\'e OS Fingerprinting (prise d'empreintes), est une \'etape cruciale 
d'un test de p\'en\'etration, puisque l'attaquant (professionnel de la s\'ecurit\'e, 
consultant ou hacker) a besoin de conna\^{\i}tre l'OS de la machine cible afin de 
choisir les exploits qu'il va utiliser. 
La d\'etection d'OS est accomplie en sniffant de fa\c{c}on passive 
des paquets r\'eseau 
et en envoyant de fa\c{c}on active des paquets de test \`a la machine cible, 
pour \'etudier des variations sp\'ecifiques dans les r\'eponses qui r\'ev\`elent 
son syst\`eme d'exploitation.

Les premi\`eres impl\'ementations de prise d'empreintes
 \'etaient bas\'ees sur l'analyse des diff\'erences entre les impl\'ementations 
de la pile TCP/IP.
 La g\'en\'eration suivante utilisa des donn\'ees de la couche d'applications,
 tels que les endpoints DCE RPC. Bien que l'analyse porte sur plus d'information,
 quelque variation de l'algorithme consistant \`a chercher le point le plus proche
 (``best fit'') est toujours utilis\'ee pour interpr\'eter cette information.
 Cette strat\'egie a plusieurs d\'efauts  : 
elle ne va pas marcher dans des situations non standard, 
et ne permet pas d'extraire les \'el\'ements clefs qui identifient de fa\c{c}on unique
 un syst\`eme d'exploitation. 
Nous pensons que le prochain pas est de travailler sur les techniques
 utilis\'ees pour analyser les donn\'ees.

Notre nouvelle approche se base sur l'analyse de la composition de
 l'information relev\'ee durant le processus d'identification du syst\`eme
 pour d\'ecouvrir les \'el\'ements clef et leurs relations.
 Pour impl\'ementer cette approche, nous avons d\'evelopp\'e des outils
 qui utilisent des r\'eseaux de neurones et des techniques des domaines
 de l'Intelligence Artificielle et les Statistiques. 
Ces outils ont \'et\'e int\'egr\'es avec succ\`es dans un 
software commercial (Core Impact).

\section{DCE-RPC Endpoint mapper}

\subsection{Le service DCE-RPC nous informe}

Dans les syst\`emes Windows, le service 
DCE (Distributed Computing Environment) RPC (Remote Procedure Call)
 re\c{c}oit les connexions envoy\'ees au port 135 de la machine cible.
 En envoyant une requ\^ete RPC, il est possible de d\'eterminer
 quels services ou programmes sont enregistr\'es dans la base de
 donn\'ees du mapper d'endpoints RPC.
 La r\'eponse inclut le UUID (Universal Unique IDentifier) 
 de chaque programme, le nom annot\'e, 
le protocole utilis\'e,
 l'adresse de r\'eseau \`a laquelle le programme est li\'e, 
et le point final du programme (endpoint).

Il est possible de distinguer les versions, \'editions et service packs
 de Windows selon la combinaison de point finaux fournie par
 le service DCE-RPC.

Par exemple, une machine Windows 2000 \'edition professionnelle
 service pack 0, le service RPC retourne 8 points finaux qui 
correspondent \`a 3 programmes :

\begin{verbatim}
uuid="5A7B91F8-FF00-11D0-A9B2-00C04FB6E6FC"
annotation="Messenger Service"
 protocol="ncalrpc"      endpoint="ntsvcs"       id="msgsvc.1" 
 protocol="ncacn_np"     endpoint="\PIPE\ntsvcs" id="msgsvc.2" 
 protocol="ncacn_np"     endpoint="\PIPE\scerpc" id="msgsvc.3" 
 protocol="ncadg_ip_udp"                         id="msgsvc.4" 

uuid="1FF70682-0A51-30E8-076D-740BE8CEE98B"
 protocol="ncalrpc"      endpoint="LRPC"         id="mstask.1" 
 protocol="ncacn_ip_tcp"                         id="mstask.2" 

uuid="378E52B0-C0A9-11CF-822D-00AA0051E40F"
 protocol="ncalrpc"      endpoint="LRPC"         id="mstask.3" 
 protocol="ncacn_ip_tcp"                         id="mstask.4" 
\end{verbatim}

\subsection{Les r\'eseaux de neurones entrent en jeu ... }

Notre id\'ee est de modeler la fonction qui fait correspondre les combinaisons de
 points finaux aux versions du syst\`eme d'exploitation avec un r\'eseau de
 neurones.

Plusieurs questions se posent :\\
$\cdot \,$ {quel genre de r\'eseau de neurones allons nous utiliser ?} \\
$\cdot \,$ {comment organiser les neurones ?} \\
$\cdot \,$ {comment faire correspondre les combinaisons de points finaux avec les
 neurones d'entr\'ee du r\'eseau ?} \\
$\cdot \,$ {comment entra\^{\i}ner le r\'eseau ?} \\

Le choix adopt\'e est d'utiliser un r\'eseau perceptron multicouches, 
plus pr\'ecis\'ement compos\'e de 3 couches 
(nous indiquons entre parenth\`eses le nombre de neurones pour 
chaque couche qui r\'esulta des tests de notre laboratoire,
voir la figure 1 pour un sch\'ema de la topologie du r\'eseau).
\begin{enumerate}
\item{couche d'entr\'ee (avec 413 neurones)
 contient un neurone pour chaque UUID
 et un neurone pour chaque point final qui correspond \`a cet UUID.
 Suivant l'exemple pr\'ec\'edent, nous avons un neurone pour le service
 Messenger et 4 neurones pour chaque endpoint associ\'e \`a ce programme. 
Cela nous permet de r\'epondre avec flexibilit\'e \`a l'apparition d'un point final
 inconnu : nous retenons de toute fa\c{c}on l'information de l'UUID principal.}
\item{couche de neurones cach\'es (avec 42 neurones),
 o\`u chaque neurone repr\'esente une combinaison des neurones d'entr\'ee.}
\item{couche de sortie (avec 25 neurones),
 contient un neurone pour chaque version et \'edition de Windows 
(p.ex. Windows 2000 \'edition professionnelle), 
et un neurone pour chaque version et service pack de Windows
 (p.ex. Windows 2000 service pack 2). De cette mani\`ere le r\'eseau peut
 distinguer l'\'edition et le service pack de fa\c{c}on ind\'ependante : 
les erreurs dans une dimension n'affectent pas les erreurs dans l'autre dimension.}
\end{enumerate}

\begin{figure}[h]
\centering
\includegraphics[width=12cm]{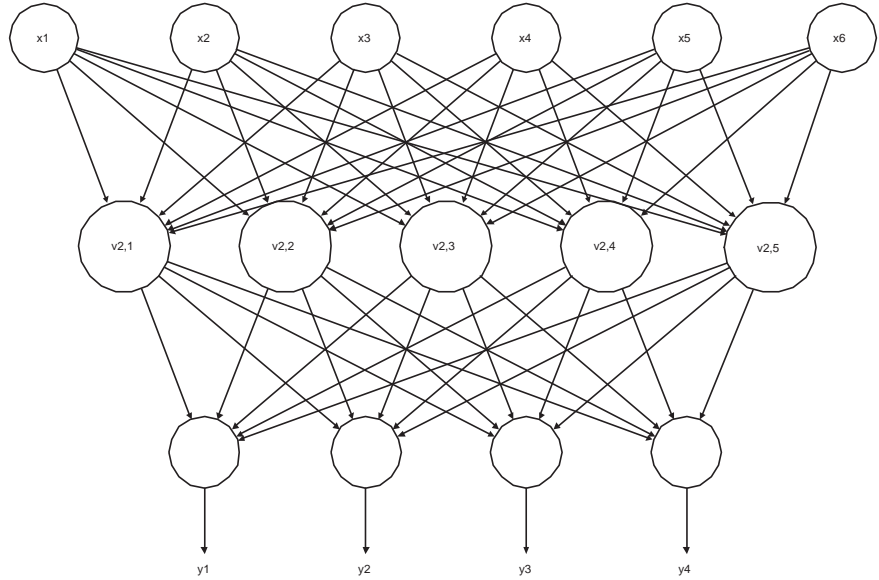}
\caption{R\'eseau de Neurones Perceptron Multicouches}
\end{figure}

Qu'est-ce qu'un perceptron ? C'est l'unit\'e de base du r\'eseau, et en suivant
 l'analogie biologique, il joue le r\^ole d'un neurone qui re\c{c}oit de l'\'energie
 \`a travers de ses connections synaptiques 
et la transmet \`a son tour selon une fonction d'activation.

Concr\`etement, chaque perceptron calcule sa valeur de sortie comme 
$$
 v_{i,j} = f ( \sum_{k = 0}^{n} \; w_{i,j,k} \cdot x_k \; ) 
$$ 
o\`u $\{ x_1 \ldots x_n \}$ sont les entr\'ees du neurone,
$x_0 = -1$ est une entr\'ee de biais fixe
et $f$ est une fonction d'activation non lin\'eaire
(nous utilisons tanh, la tangente hyperbolique).
L'entra\^{\i}nement du r\'eseau consiste \`a calculer les poids synaptiques
$ \{ w_{i,j,0} \ldots w_{i,j,n} \} $ 
pour chaque neurone $(i,j)$ ($i$ est la couche,
$j$ est la position du neurone dans la couche).

\subsection{Entra\^{\i}nement par r\'etropropagation}

L'entra\^{\i}nement se r\'ealise par r\'etropropagation :
pour la couche de sortie, \`a partir d'une sortie attendue
$ \{ y_1 \ldots y_m \} $ 
nous calculons (pour chaque neurone) une estimation de l'erreur
$$
\delta_{i,j} = f'(v_{i,j}) \; ( y_j - v_{i,j} )
$$
Celle-ci est propag\'ee aux couches pr\'ec\'edentes par :
$$
\delta_{i,j} = f'(v_{i,j}) \; \sum_k w_{i,j,k} \cdot \delta_{i+1, j}
$$
o\`u la somme est calcul\'ee sur tous les neurones connect\'es
avec le neurone $(i,j)$.

Les nouveaux poids, au temps $t+1$, sont
$$
w_{t+1; i,j,k} = w_{t; i,j,k} + \Delta w_{t; i,j,k} 
$$ 
o\`u $\Delta w_t$ d\'epend d'un facteur de correction
et aussi de la valeur de 
$\Delta w_{t-1}$ 
multipli\'ee por un momentum $\mu$ 
(cela donne aux modifications une sorte d'\'energie cin\'etique) :
$$
\Delta w_{t; i,j,k} = ( \lambda \cdot \delta_{i+1,k} \cdot v_{i,j}  ) + \mu \cdot \Delta w_{t-1; i,j,k} 
$$
Le facteur de correction d\'epend des valeurs $\delta$ calcul\'ees
et aussi d'un facteur d'apprentissage $\lambda$ qui peut \^etre ajust\'e
pour acc\'el\'erer la convergence du r\'eseau.

Le type d'entra\^{\i}nement r\'ealis\'e s'appelle apprentissage supervis\'e
(bas\'e sur des exemples).
La r\'etropropagation part d'un jeu de donn\'ees 
avec des entr\'ees et des sorties recherch\'ees.
Une g\'en\'eration consiste \`a recalculer les poids synaptiques 
pour chaque paire  d'entr\'ee / sortie. 
L'entra\^{\i}nement complet requiert 10350 g\'en\'erations, 
ce qui prend quelques 14 heures d'ex\'ecution de code python.
Etant donn\'e que le design de la topologie du r\'eseau est un 
processus assez artisanal (essai et erreur),
qui requiert d'entra\^{\i}ner le r\'eseau pour chaque variation de 
la topologie afin de voir si elle produit de meilleurs r\'esultats,
le temps d' ex\'ecution nous a particuli\`erement motiv\'es
\`a am\'eliorer la vitesse de convergence de l'entra\^{\i}nement
(probl\`eme que nous traitons dans la section 4).

Pour chaque g\'en\'eration du processus d'entra\^{\i}nement,
 les entr\'ees sont r\'eordonn\'ees au hasard 
(pour que l'ordre des donn\'ees n'affecte pas l'entra\^{\i}nement). 
En effet le r\'eseau est susceptible d'apprendre
n'importe quel aspect des entr\'ees fournies, 
en particulier l'ordre dans lequel on lui pr\'esente les choses.  

La table \ref{tab-comparaison} montre une comparaison (de notre laboratoire) entre
le vieux module DCE-RPC (qui utilise un algorithme de ``best fit'')
et le nouveau module qui utilise un r\'eseau de neurones
 pour analyser l'information.

\begin{table} [h]
\centering
\begin{tabular} {| l | c | c | }
\hline
R\'esultat & Ancien module & DCE-RPC avec \\
 & DCE-RPC & r\'eseau de neurones \\
\hline
Concordance parfaite & 6 & 7 \\
Concordance partielle & 8 & 14 \\
Erreur & 7 & 0 \\
Pas de r\'eponse & 2 & 2 \\
\hline
\end{tabular}
\vskip 0.3 cm
\caption{Comparaison entre l'ancien module DCE-RPC et le nouveau module}
\label{tab-comparaison}
\end{table}

La table \ref{tab-resultat} montre le r\'esultat de l'ex\'ecution du module contre un 
Windows 2000 \'edition server sp1. 
Le syst\`eme correct est reconnu avec un haut niveau de confiance.

\begin{table} [p]
\begin{verbatim}
Neural Network Output (close to 1 is better):
Windows NT4: 4.87480503763e-005
Editions:
    Enterprise Server: 0.00972694324639
    Server: -0.00963500026763
Service Packs:
    6: 0.00559659167371
    6a: -0.00846224120952
Windows 2000: 0.996048928128
Editions:
    Server: 0.977780526016
    Professional: 0.00868998746624
    Advanced Server: -0.00564873813703
Service Packs:
    4: -0.00505441088081
    2: -0.00285674134367
    3: -0.0093665583402
    0: -0.00320117552666
    1: 0.921351036343
Windows 2003: 0.00302898647853
Editions:
    Web Edition: 0.00128127138728
    Enterprise Edition: 0.00771786077082
    Standard Edition: -0.0077145024893
Service Packs:
    0: 0.000853988551952
Windows XP: 0.00605168045887
Editions:
    Professional: 0.00115635710749
    Home: 0.000408057333416
Service Packs:
    2: -0.00160404945542
    0: 0.00216065240615
    1: 0.000759109188052
Setting OS to Windows 2000 Server sp1
Setting architecture: i386
\end{verbatim}
\caption{R\'esultat du module DCE-RPC endpoint mapper}
\label{tab-resultat}
\end{table}

\section{D\'etection d'OS bas\'ee sur les signatures de Nmap}

\subsection{Richesse et faiblesse de Nmap}

Nmap est un outil d'exploration r\'eseau et un scanner de s\'ecurit\'e qui
inclut une m\'ethode de d\'etection d'OS
bas\'ee sur la r\'eponse de la machine cible \`a une s\'erie de 9 tests.
Nous d\'ecrivons bri\`evement dans le tableau \ref{tests-nmap} les paquets envoy\'es 
par chaque test, pour plus d'informations voir la page de Nmap \cite{nmap}.

\begin{table}
\centering
\begin{tabular}{| l | l | l | l  |}
\hline
Test  & envoyer paquet   &  \`a un port  &  avec les flags \\
\hline
T1  &  TCP  &  TCP ouvert &  SYN, ECN-Echo  \\
T2  &  TCP  &  TCP ouvert  &  no flags \\
T3  &  TCP  &  TCP ouvert &  URG, PSH, SYN, FIN \\
T4  &  TCP  &  TCP ouvert &  ACK \\
T5  &  TCP  &  TCP ferm\'e &  SYN \\
T6  &  TCP  &  TCP ferm\'e &  ACK \\
T7  &  TCP  &  TCP ferm\'e &  URG, PSH, FIN \\
PU &  UDP  &  UDP ferm\'e &  \\
TSeq &  TCP * 6  & TCP ouvert & SYN \\
\hline
\end{tabular}
\vskip 0.3 cm
\caption{Description des paquets envoy\'es}
\label{tests-nmap}
\end{table}

Notre m\'ethode utilise la base de signatures de Nmap.
Une signature est un ensemble de r\`egles d\'ecrivant comment 
une version / \'edition sp\'ecifique d'un syst\`eme d'exploitation 
r\'epond aux tests. 
Par exemple :

\begin{verbatim}
# Linux 2.6.0-test5 x86
Fingerprint Linux 2.6.0-test5 x86
Class Linux | Linux | 2.6.X | general purpose
TSeq(Class=RI%gcd=<6%SI=<2D3CFA0&>73C6B%IPID=Z%TS=1000HZ)
T1(DF=Y%W=16A0%ACK=S++%Flags=AS%Ops=MNNTNW)
T2(Resp=Y%DF=Y%W=0%ACK=S%Flags=AR%Ops=)
T3(Resp=Y%DF=Y%W=16A0%ACK=S++%Flags=AS%Ops=MNNTNW)
T4(DF=Y%W=0%ACK=O%Flags=R%Ops=)
T5(DF=Y%W=0%ACK=S++%Flags=AR%Ops=)
T6(DF=Y%W=0%ACK=O%Flags=R%Ops=)
T7(DF=Y%W=0%ACK=S++%Flags=AR%Ops=)
PU(DF=N%TOS=C0%IPLEN=164%RIPTL=148%RID=E%RIPCK=E%UCK=E%ULEN=134%DAT=E)
\end{verbatim}

La base de Nmap contient 1684 signatures, ce qui veut dire que 
quelques 1684 versions / \'editions de syst\`emes d'exploitation 
peuvent th\'eoriquement \^etre distingu\'ees par cette m\'ethode.

Nmap fonctionne en comparant la r\'eponse d'une machine avec 
chaque signature de la base de donn\'ees.
Un score est assign\'e \`a chaque signature, calcul\'e simplement
comme le nombre de r\`egles qui concordent divis\'e par 
le nombre de r\`egles consid\'er\'ees 
(en effet, les signatures peuvent avoir diff\'erents nombres de r\`egles,
ou quelques champs peuvent manquer dans la r\'eponse,
dans ce cas les r\`egles correspondantes ne sont pas prises en compte).
C'est-\`a-dire que Nmap effectue une esp\`ece de 
``best fit'' bas\'e sur une distance de Hamming,
o\`u tous les champs de la r\'eponse ont le m\^eme poids.

Un des probl\`emes que pr\'esente cette m\'ethode est le suivant :
 les syst\`emes d'exploitation rares (improbables)
qui g\'en\`erent moins de r\'eponses aux tests
obtiennent un meilleur score ! 
(les r\`egles qui concordent acqui\`erent un plus grand poids relatif).  
Par exemple, il arrive que Nmap d\'etecte un Windows 2000
comme un Atari 2600 ou un HPUX ...
La richesse de la base de donn\'ees devient alors une faiblesse !

\subsection{Structure de R\'eseaux Hi\'erarchique}

Si nous repr\'esentons symboliquement l'espace des syst\`emes d'exploitation
comme un espace \`a 568 dimensions 
(nous verrons par la suite le pourquoi de ce nombre),
les r\'eponses possibles des diff\'erentes versions des syst\`emes inclus dans
la base de donn\'ees forme un nuage de points.
Ce grand nuage est structur\'e de fa\c{c}on particuli\`ere,
puisque les familles de syst\`emes d'exploitation forment
des clusters plus ou moins reconnaissables.
La m\'ethode de Nmap consiste, \`a partir de la r\'eponse d'une machine,
\`a chercher le point le plus proche 
(selon la distance de Hamming d\'ej\`a mentionn\'ee).

Notre approche consiste en premier lieu \`a filtrer les
syst\`emes d'exploitation qui ne nous int\'eressent pas
(toujours selon le point de vue de l'attaquant, 
par exemple les syst\`emes pour lesquels il n'a pas d'exploits).
Dans notre impl\'ementation, nous sommes int\'eress\'es par
les familles Windows, Linux, Solaris, OpenBSD, NetBSD et FreeBSD.
Ensuite nous utilisons la structure des familles de syst\`emes d'exploitation
pour assigner la machine \`a une des 6 familles consid\'er\'ees.

Le r\'esultat est un module qui utilise plusieurs r\'eseaux de neurones
organis\'es de fa\c{c}on hi\'erarchique :
\begin{enumerate}
\item{
premier pas, un r\'eseau de neurones pour d\'ecider si l'OS est int\'eressant ou non.}
\item{
deuxi\`eme pas, un r\'eseau de neurones pour d\'ecider la famille de l'OS :
Windows, Linux, Solaris, OpenBSD, FreeBSD, NetBSD.}
\item{
dans le cas de Windows, nous utilisons le module
DCE-RPC endpoint mapper pour raffiner la d\'etection.}
\item{
dans le cas de Linux, nous r\'ealisons une analyse conditionn\'ee 
(avec un  autre r\'eseau de neurones)
pour d\'ecider la version du kernel.}
\item{
dans le cas de Solaris et des BSD, nous r\'ealisons une analyse 
conditionn\'ee pour d\'ecider la version.}
\end{enumerate}

Nous utilisons un r\'eseau de neurones diff\'erent pour chaque analyse.
Nous avons ainsi 5 r\'eseaux de neurones, et chacun requiert
une topologie et un entra\^{\i}nement sp\'ecial.

\subsection{Entr\'ees du r\'eseau de neurones}

La premi\`ere question \`a r\'esoudre est :
comment traduire la r\'eponse d'une machine
en entr\'ees pour le r\'eseau de neurones ?
Nous assignons un ensemble de neurones d'entr\'ee \`a chaque test.

\noindent
Voici les d\'etails pour les tests T1 ... T7 :\\
$\cdot \,$ un neurone pour le flag ACK.\\
$\cdot \,$ un neurone pour chaque r\'eponse : S, S++, O.\\
$\cdot \,$ un neurone pour le flag DF.\\
$\cdot \,$ un neurone pour la r\'eponse : yes/no.\\
$\cdot \,$ un neurone pour le champ \emph{Flags}.\\
$\cdot \,$ un neurone pour chaque flag : ECE, URG, ACK, PSH, RST, SYN, FIN
(en total 8 neurones).\\
$\cdot \,$ 10 groupes de 6 neurones pour le champ \emph{Options}.
Nous activons un seul neurone dans chaque groupe suivant l'option -
EOL, MAXSEG, NOP, TIMESTAMP, WINDOW, ECHOED -
en respectant l'ordre d'apparition (soit au total 60 neurones pour les options). \\
$\cdot \,$ un neurone pour le champ $W$ (window size), qui a pour entr\'ee
une valeur hexad\'ecimale.

Pour les flags ou les options, l'entr\'ee est 1 ou -1 (pr\'esent ou absent).
D'autres neurones ont une entr\'ee num\'erique, comme
le champ W (window size),
le champ GCD (plus grand commun diviseur des num\'eros de s\'equence initiaux)
ou les champs SI et VAL des r\'eponses au test Tseq.
Dans l'exemple d'un Linux 2.6.0, la r\'eponse  :
\begin{verbatim} 
T3(Resp=Y%DF=Y%W=16A0%ACK=S++%Flags=AS%Ops=MNNTNW)
\end{verbatim}

\begin{table}
\centering
\begin{tabular} {| c | c | c | c | c |  c | c | c | c | c |  c | c | c | c | c |    }
\hline
ACK & S & S++ & O & DF & Yes & Flags & E & U & A & P & R & S & F & $\ldots$ \\
\hline
1     & -1 & 1     & -1 &  1  &   1   & 1        & -1    &  -1     &   1    &   -1    &  -1   &   1   &  -1   &  $\ldots$ \\
\hline
\end{tabular}
\vskip 0.3 cm
\caption{R\'esultat de la transformation (Linux 2.6.0)}
\label{tab-resultat-linux}
\end{table}

\noindent
se transforme en comme indiqu\'e dans le tableau \ref{tab-resultat-linux}.
De cette fa\c{c}on nous obtenons une couche d'entr\'ee avec 568 dimensions,
avec une certaine redondance.
La redondance nous permet de traiter de fa\c{c}on flexible les r\'eponses
 inconnues mais introduit aussi des probl\`emes de performance !
Nous verrons dans la section suivante comment r\'esoudre ce probl\`eme
(en r\'eduisant le nombre de dimensions).
Comme pour le module DCE-RPC, 
les r\'eseaux de neurones sont compos\'es de 3 couches.
Par exemple le premier r\'eseaux de neurones (le filtre de pertinence) contient :
la couche d'entr\'ee 96 neurones,
la couche cach\'ee 20 neurones,
la couche de sortie 1 neurone.

\subsection{G\'en\'eration du jeu de donn\'ees}

Pour entra\^{\i}ner le r\'eseau de neurones nous avons besoin
d'entr\'ees (r\'eponses de machines)
avec les sorties correspondantes (OS de la machine).
Comme la base de signatures contient 1684 r\`egles,
nous estimons qu'une population de 15000 machines
est n\'ecessaire pour entra\^{\i}ner le r\'eseau.
Nous n'avons pas acc\`es \`a une telle population ...
et scanner l'Internet n'est pas une option !

La solution que nous avons adopt\'ee est de g\'en\'erer les entr\'ees 
par une simulation Monte Carlo.
Pour chaque r\`egle, nous g\'en\'erons des entr\'ees correspondant
 \`a cette r\`egle.
Le nombre d'entr\'ees d\'epend de la distribution empirique des 
OS bas\'ee sur des donn\'ees statistiques.
Quand la r\`egle sp\'ecifie une constante, nous utilisons cette valeur,
et quand la r\`egle sp\'ecifie des options ou un intervalle de valeurs,
nous choisissons une valeur en suivant une distribution uniforme.

\section{R\'eduction des dimensions et entra\^{\i}nement}

\subsection{Matrice de corr\'elation}

Lors du design de la topologie des r\'eseaux, nous avons \'et\'e
 g\'en\'ereux avec les entr\'ees  :
568 dimensions, avec une redondance importante.
Une cons\'equence est que la convergence de l'entra\^{\i}nement est lente,
d'autant plus que le jeu de donn\'ees est tr\`es grand.
La solution \`a ce probl\`eme fut de r\'eduire le nombre de dimensions.
Cette analyse nous permet aussi de mieux comprendre
les \'el\'ements importants des tests utilis\'es.

Le premier pas est de 
consid\'erer chaque dimension d'entr\'ee comme une variable al\'eatoire
$ X_i \, ( 1 \leq i \leq 568) $.
Les dimensions d'entr\'ee ont des ordres de grandeur diff\'erents  :
les flags prennent comme valeur 1/-1 alors que
le champ ISN (num\'ero de s\'equence initial) est un entier de 32 bits.
Nous \'evitons au r\'eseau de neurones d'avoir \`a apprendre
\`a additionner correctement ces variables h\'et\'erog\`enes 
 en normalisant les variables al\'eatoires
(en soustrayant la moyenne $\mu$
et en divisant par l'\'ecart type $\sigma$) :
$$
\frac { X_i - \mu_i } { \sigma_i }
$$
Puis nous calculons la matrice de corr\'elation $R$,
dont les \'el\'ements sont :
$$
R_{i,j} = \frac{ E[ (X_i - \mu_i ) (X_j - \mu_j) ] } { \sigma_i \; \sigma_j }
$$
Le symbole $E$ d\'esigne l'esp\'erance math\'ematique.
Puisqu'apr\`es avoir normalis\'e les variables,
$\mu_i = 0$ et $\sigma_i = 1$ pour tout $i$, 
la matrice de corr\'elation est simplement
$ R_{i,j} = E[ X_i \,X_j ] $.

La corr\'elation est une mesure de la d\'ependance statistique
entre deux variables
(une valeur proche de 1 ou -1 indique une plus forte d\'ependance).
La d\'ependance lin\'eaire entre des colonnes de $R$ indique 
des variables d\'ependantes,
dans ce cas nous en gardons une et \'eliminons les autres,
puisqu'elles n'apportent pas d'information additionnelle.
Les constantes ont une variance nulle et sont aussi \'elimin\'ees
par cette analyse.

Voyons le r\'esultat dans le cas des syst\`emes OpenBSD. 
Nous reproduisons ci-dessous les extraits des
signatures de deux OpenBSD diff\'erents,
o\`u les champs qui survivent \`a la r\'eduction de la 
matrice de corr\'elation sont marqu\'es en italiques.

\noindent \texttt{Fingerprint OpenBSD 3.6 (i386) \\
Class OpenBSD | OpenBSD | 3.X | general purpose \\
T1(DF=N \% \emph{W=4000} \% ACK=S++ \% Flags=AS \% Ops=\emph{MN}WNNT) \\
T2(Resp=N) \\
T3(\emph{Resp=N}) \\
T4(DF=N \% \emph{W=0} \% ACK=O \%Flags=R \% Ops=)   \\
T5(DF=N \% W=0 \% ACK=S++ \% Flags=AR \% Ops=)  }

\noindent \texttt{Fingerprint OpenBSD 2.2 - 2.3 \\
Class OpenBSD | OpenBSD | 2.X | general purpose \\
T1(DF=N \% \emph{W=402E} \% ACK=S++ \% Flags=AS \% Ops=\emph{MN}WNNT) \\
T2(Resp=N) \\
T3(\emph{Resp=Y} \% DF=N \% \emph{W=402E} \% ACK=S++ \% Flags=AS \% Ops=\emph{MN}WNNT) \\
T4(DF=N \% \emph{W=4000} \% ACK=O \% Flags=R \% Ops=) \\
T5(DF=N \% W=0 \% ACK=S++ \% Flags=AR \% Ops=)
}

Par exemple, pour le test T1, les seuls champs qui varient
sont $W$ et les deux premi\`eres options, 
les autres sont constants dans toutes les versions de OpenBSD.
Autre exemple, pour le test T4 seul $W$ est susceptible de varier,
et le test T5 n'apporte directement aucune information
sur la version de OpenBSD examin\'ee.

\begin{table}
\centering
\begin{tabular} { | c | c | l | }
\hline
Index & Ancien index & Nom du champ \\
\hline
0 &   20  &  T1 : TCP OPT 1 EOL \\
1 &   26  &  T1 : TCP OPT 2 EOL \\
2  &  29  &  T1 : TCP OPT 2 TIMESTAMP \\
3  &  74  &  T1 : W FIELD \\
4  &  75  &  T2 : ACK FIELD \\
5  &  149  &  T2 : W FIELD \\
6 &   150  &  T3 : ACK FIELD \\
7  &  170  &  T3 : TCP OPT 1 EOL \\
8  &  179  &  T3 : TCP OPT 2 TIMESTAMP \\
9  &  224  &  T3 : W FIELD \\
10  &  227  &  T4 : SEQ S \\
11  &  299  &  T4 : W FIELD \\
12  &  377  &  T6 : SEQ S \\
13  &  452  &  T7 :  SEQ S \\
14  &  525  &  TSeq : CLASS FIELD \\
15  &  526  &  TSeq : SEQ TD \\
16  &  528  &  TSeq : SEQ RI \\
17  &  529  &  TSeq : SEQ TR \\
18  &  532  &  TSeq : GCD FIELD \\
19  &  533  &  TSeq : IPID FIELD \\
20  &  535  &  TSeq : IPID SEQ BROKEN INCR \\
21  &  536  &  TSeq : IPID SEQ RPI \\
22  &  537  &  TSeq : IPID SEQ RD \\
23  &  540  &  TSeq : SI FIELD \\
24  &  543  &  TSeq : TS SEQ 2HZ \\
25  &  546  &  TSeq : TS SEQ UNSUPPORTED \\
26  &  555  &  PU : UCK RID RIPCK EQ \\
27  &  558  &  PU : UCK RID RIPCK ZERO \\
28  &  559  &  PU : UCK RID RIPCK EQ \\
29  &  560  &  PU : UCK RID RIPCK FAIL \\
30  &  563  &  PU : UCK RID RIPCK EQ \\
31  &  564  &  PU : UCK RID RIPCK FAIL \\
32  &  565  &  PU : RIPTL FIELD \\
33  &  566  &  PU : TOS FIELD \\
\hline
\end{tabular}
\vskip 0.3 cm
\caption{Champs qui permettent de distinguer les OpenBSD}
\label{tab-openbsd}
\end{table}

La table \ref{tab-openbsd} montre la liste compl\`ete des champs qui servent \`a
distinguer les diff\'erentes versions d'OpenBSD.
Comme nous l'avons dit, le test T5 n'appara\^{\i}t pas,
alors que les tests Tseq et PU conservent de nombreuses
variables, ce qui nous montre que
ces deux tests sont les plus discriminatifs
au sein de la population OpenBSD.

\subsection{Analyse en Composantes Principales}

Une r\'eduction ult\'erieure des donn\'ees utilise l'Analyse en Composantes Principales (ACP).
L'id\'ee est de calculer une nouvelle base (ou syst\`eme de coordonn\'ees) 
de l'espace d'entr\'ee, de telle mani\`ere que
la majeure variance de toute projection du jeu de donn\'ees dans 
un sous-espace de  $k$  dimensions,
provient de projeter sur les  $k$  premiers vecteurs de cette base.

\noindent
L'algorithme ACP consiste \`a  : \\
$\cdot \,$ calculer les vecteurs propres et valeurs propres de $R$.\\ 
$\cdot \,$ trier les vecteurs par valeur propre d\'ecroissante.\\
$\cdot \,$ garder les  $k$  premiers vecteurs pour projeter les donn\'ees.\\
$\cdot \,$ le param\`etre  $k$  est choisi pour maintenir 
au moins 98\% de la variance totale.

Apr\`es avoir r\'ealis\'e l'ACP nous avons obtenu les topologies indiqu\'ee dans le tableau \ref{tab-topologies}
pour les r\'eseaux de neurones 
(la taille de la couche d'entr\'ee originale \'etait de 568 dans tous les cas).

\begin{table}
\centering
\begin{tabular} { | l | c | c | c | c | }
\hline
Analyse &  Couche d'entr\'ee & Couche d'entr\'ee 
& Couche & Couche  \\
 & (apr\`es r\'eduction& (apr\`es ACP) & cach\'ee & de sortie\\
& de la matrice $R$) & & & \\
\hline
Pertinence & 204 & 96 & 20 & 1 \\
Famille d'OS & 145 & 66 & 20 & 6 \\
Linux & 100 & 41 & 18 & 8 \\
Solaris & 55 & 26 & 7 & 5 \\
OpenBSD & 34 & 23 & 4 & 3 \\
\hline
\end{tabular}
\vskip 0.3 cm
\caption{Topologies des r\'eseaux de neurones}
\label{tab-topologies}
\end{table}

Pour conclure l'exemple de OpenBSD, \`a partir des 34 variables qui
ont surv\'ecu \`a la r\'eduction de la matrice de corr\'elation,
il est possible de construire une nouvelle base de 23 vecteurs.
Les coordonn\'ees dans cette base sont les entr\'ees du r\'eseau, 
la couche cach\'ee ne contient que 4 neurones et la couche de
sortie 3 neurones (car nous distinguons 3 groupes de versions
d'OpenBSD). 
Une fois que l'on sait qu'une machine est un OpenBSD, 
le probl\`eme de reconna\^{\i}tre la version est beaucoup plus simple
et born\'e, et peut \^etre accompli par un r\'eseau de neurones de petite
taille (plus efficient et rapide).

\subsection{Taux d'apprentissage adaptatif}

C'est une strat\'egie pour acc\'el\'erer la convergence de l'entra\^{\i}nement.
Le taux d'apprentissage est le param\`etre $\lambda$ qui intervient
dans les formules d'apprentissage par r\'etropropagation.

Etant donn\'ee une sortie du r\'eseau, nous pouvons
calculer une estimation de l'erreur quadratique
$$
\frac { \sum_{i=1}^{n} ( y_i - v_i ) ^ 2 } {n}
$$
o\`u  $y_i$  sont les sorties recherch\'ees et  $v_i$  sont les sorties du r\'eseau.

Apr\`es chaque g\'en\'eration (c'est-\`a-dire apr\`es avoir fait les calculs 
pour toutes les paires d'entr\'ee / sortie),
si l'erreur est plus grande, nous diminuons le taux d'apprentissage.
Au contraire, si l'erreur est plus petite, alors nous augmentons le taux d'apprentissage.
L'id\'ee est de se d\'eplacer plus rapidement si nous allons dans la direction correcte.

\begin{figure}[t]
\centering
\includegraphics[width=12cm]{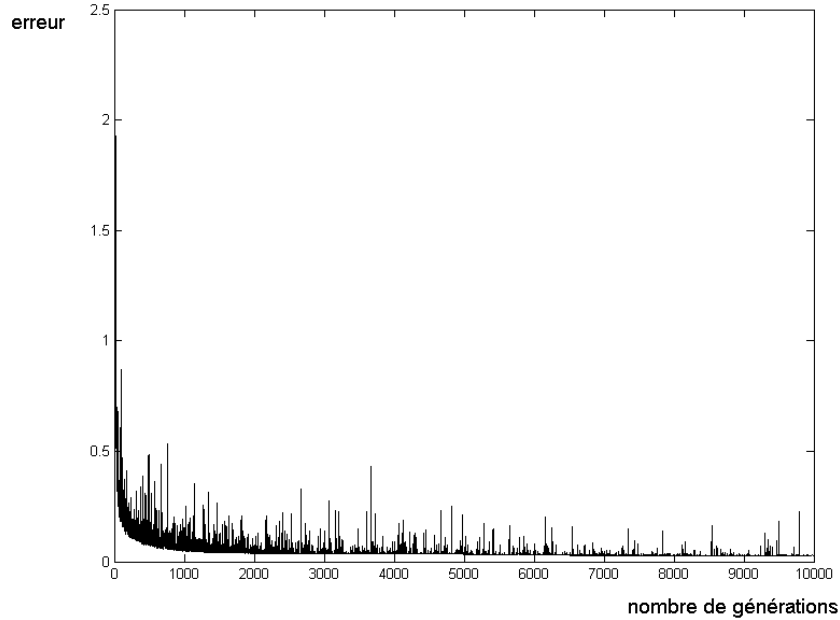}
\caption{Taux d'apprentissage fixe}
\end{figure}

\begin{figure}[t]
\centering
\includegraphics[width=12cm]{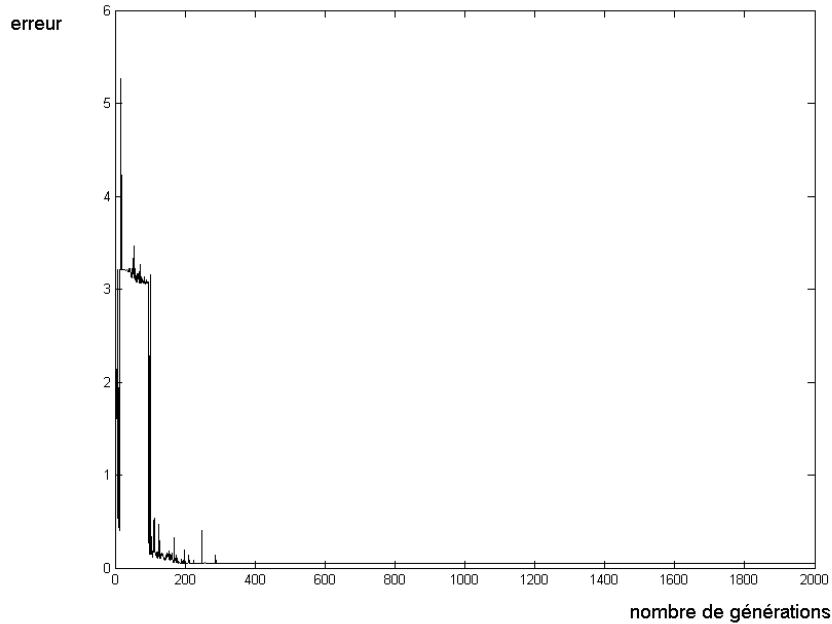}
\caption{Taux d'apprentissage adaptatif}
\end{figure}

Voici deux graphiques (Figures 2 et 3) qui montrent l'\'evolution de l'erreur quadratique
moyenne en fonction du nombre de g\'en\'erations pour chaque
strat\'egie.
Lorsque le taux d'apprentissage est fixe,
l'erreur diminue et atteint des valeurs satisfaisantes apr\`es 4000 ou 5000
g\'en\'erations. L'erreur a une claire tendance \`a la baisse
(les r\'esultats sont bons) mais avec des pics irr\'eguliers,
dus \`a la nature probabiliste de l'entra\^{\i}nement du r\'eseau.

En utilisant un taux d'apprentissage adaptatif, 
nous obtenons au d\'ebut un comportement plus chaotique,
avec des niveaux d'erreur plus hauts. 
Mais une fois que le syst\`eme trouve la direction correcte,
l'erreur chute rapidement pour atteindre une valeur 
tr\`es faible et constante apr\`es 400 g\'en\'erations. 
Ces r\'esultats sont clairement meilleurs et permettent 
d'acc\'el\'erer l'entra\^{\i}nement des r\'eseaux.

\subsection{Entra\^{\i}nement par sous-ensembles du jeu de donn\'ees}

C'est une autre strat\'egie pour acc\'el\'erer la convergence de l'entra\^{\i}nement.
Elle consiste \`a entra\^{\i}ner le r\'eseau avec plusieurs sous-ensembles 
plus petits du jeu de donn\'ees.
Ceci permet aussi de r\'esoudre des probl\`emes de limitation
d'espace, par exemple lorsque le jeu de donn\'ees est trop grand
pour \^etre charg\'e entier en m\'emoire.

Pour estimer l'erreur commise apr\`es avoir utilis\'e un sous-ensemble,
 nous calculons une mesure d'ad\'equation $G$.
Si la sortie est 0/1 : 
$$
		G = 1 - ( \Pr[ \mbox{ faux positif } ] + \Pr[ \mbox{ faux n\'egatif } ] )
$$
Pour d'autres type de sorties, $G$ est simplement : 
$$
		G = 1 - {\mbox{nombre d'erreurs}} / {\mbox{nombre de sorties}}.
$$
A nouveau, nous utilisons une strat\'egie de 
taux d'apprentissage adaptatif.
Si la mesure d'ad\'equation $G$ augmente, 
alors nous augmentons le taux d'apprentissage initial 
(pour chaque sous-ensemble).

Nous reproduisons ci-dessous le r\'esultat de l'ex\'ecution du module
contre une machine Solaris 8. 
Le syst\`eme correct est reconnu avec pr\'ecision.
\begin{verbatim}
Relevant / not relevant analysis
    0.99999999999999789    relevant 

Operating System analysis
    -0.99999999999999434   Linux 
    0.99999999921394744    Solaris 
    -0.99999999999998057   OpenBSD
    -0.99999964651426454 	 FreeBSD 
    -1.0000000000000000    NetBSD
    -1.0000000000000000    Windows

Solaris version analysis
    0.98172780325074482    Solaris 8 
    -0.99281382458335776   Solaris 9 
    -0.99357586906143880   Solaris 7 
    -0.99988378968003799   Solaris 2.X 
    -0.99999999977837983   Solaris 2.5.X 
\end{verbatim}

\section{Conclusion et id\'ees pour le futur}

Dans ce travail, nous avons vu que l'une des principales limitations
des techniques classiques de d\'etection des
syst\`emes d'exploitation r\'eside dans l'analyse des donn\'ees
recueillies par les tests, bas\'ee sur quelque variation de 
l'algorithme de ``best fit'' (chercher le point le plus proche en 
fonction d'une distance de Hamming).

Nous avons vu comment g\'en\'erer et r\'eunir l'information \`a analyser,
comment homog\'en\'eiser les donn\'ees 
(normaliser les variables d'entr\'ee) 
et surtout comment d\'egager la  structure des donn\'ees d'entr\'ee.
C'est l'id\'ee principale de notre approche, qui motive la d\'ecision
d'utiliser des r\'eseaux de neurones, 
de diviser l'analyse en plusieurs \'etapes hi\'erarchiques,
et de r\'eduire le nombre de dimensions d'entr\'ee.
Les r\'esultats exp\'erimentaux (de notre laboratoire)
montrent que cette approche permet d'obtenir une
reconnaissance plus fiable des syst\`emes d'exploitation.

De plus, la r\'eduction de la matrice de corr\'elation 
et l'analyse en composantes principales,
introduits en principe pour r\'eduire le nombre de dimensions
et am\'eliorer la convergence de l'entra\^{\i}nement,
nous donnent une m\'ethode syst\'ematique pour analyser
les r\'eponses des machines aux stimuli envoy\'es.
Cela nous a permis de d\'egager les \'el\'ements clefs des tests de Nmap,
voir par exemple la table des champs permettant de
distinguer les diff\'erentes versions d'OpenBSD.
Une application de cette analyse serait d'optimiser
les tests de Nmap pour g\'en\'erer moins de trafic.
Une autre application plus ambitieuse serait
de cr\'eer une base de donn\'ees avec les r\'eponses d'une population 
repr\'esentative de machines \`a une vaste batterie de tests
(combinaisons de diff\'erents types de paquets, ports et flags).
 Les m\^emes m\'ethodes d'analyse permettraient de d\'egager
de cette vaste base de donn\'ees les tests les plus discriminatifs
pour la reconnaissance d'OS.

L'analyse que nous proposons peut aussi s'appliquer 
\`a d'autres m\'ethodes de d\'etection : 
\begin{enumerate}
\item{Xprobe2, d'Ofir Arkin, Fyodor \& Meder Kydyraliev, 
qui base la d\'etection sur des tests ICMP, SMB, SNMP.}

\item{Passive OS Identification (p0f) de Michal Zalewski,
m\'ethode qui a l'avantage de ne pas g\'en\'erer de trafic additionnel.
C'est un d\'efi int\'eressant, car l'analyse porte sur un 
volume de donn\'ees plus important (tout le trafic sniff\'e),
et requiert sans doute des m\'ethodes plus dynamiques 
et \'evolutives.}

\item{D\'etection d'OS bas\'ee sur l'information fournie
par le portmapper SUN RPC, permettant de distinguer
des syst\`emes Sun, Linux et autres versions de System V.}

\item{R\'eunion d'information pour le versant client-side des tests de p\'en\'etration, en particulier pour d\'etecter les versions d'applications.
Par exemple d\'etecter les 
Mail User Agents (MUA) tels que Outlook ou Thunderbird, 
en utilisant les Mail Headers.}
\end{enumerate}

Une autre id\'ee pour le futur est d'ajouter du bruit et le filtre 
d'un firewall aux donn\'ees \'etudi\'ees.
Ceci permettrait de d\'etecter la pr\'esence d'un firewall,
d'identifier diff\'erents firewalls
et de faire des tests plus robustes.


\end{document}